\documentclass[conference]{IEEEtran}
\IEEEoverridecommandlockouts
\usepackage[T1]{fontenc}
\usepackage[utf8]{inputenc}
\usepackage{cite}
\usepackage{graphicx}
\usepackage{amsmath}
\usepackage{amssymb}
\usepackage{booktabs}
\usepackage{array}
\usepackage{multirow}
\usepackage{xcolor}
\usepackage{listings}
\usepackage{tikz}
\usetikzlibrary{shapes.geometric, arrows, positioning, calc}
\usepackage{url}
\usepackage[hidelinks]{hyperref}

\newcolumntype{L}[1]{>{\raggedright\arraybackslash}p{#1}}

\definecolor{secugray}{gray}{0.45}
\lstdefinestyle{ada}{
  language=Ada, basicstyle=\ttfamily\scriptsize, keywordstyle=\bfseries,
  commentstyle=\color{secugray}\itshape, columns=fullflexible,
  showstringspaces=false, frame=tb, framerule=0.3pt, rulecolor=\color{secugray},
  aboveskip=6pt, belowskip=2pt, xleftmargin=4pt, breaklines=true}
\newcommand{\VCtotal}{49{,}280}

\newcommand{\TestTotal}{1{,}600}
\newcommand{\VCcrypto}{34{,}820}
\newcommand{\VCmatrix}{3{,}496}
\newcommand{\VCikev}{2{,}039}
\newcommand{\VCca}{2{,}696}
\newcommand{\VChashmap}{5{,}039}
\newcommand{\EffortTotalH}{16}
\newcommand{\EffortTotalEUR}{150}
\newcommand{\SpeedupLow}{20}
\newcommand{\SpeedupHigh}{40}

\begin{document}

\title{The Prover Is the Judge: Verified Security Software from AI Coding Agents in Ada/SPARK}

\author{\IEEEauthorblockN{Tobias Philipp}
\IEEEauthorblockA{\textit{secunet Security Networks AG}, Essen, Germany \\
tobias.philipp@secunet.com}}

\maketitle

\begin{abstract}
  AI coding agents produce code faster than humans can review it. In our approach, the
  prover is the judge of whether the code is correct. Under a verifier-driven loop, AI agents wrote and
  verified bare-metal security software in Ada/SPARK spanning classical and post-quantum cryptography,
  TLS\,1.3, IKEv2, X.509, and a Matrix client. GNATprove discharged \VCtotal{} proof obligations,
  established functional correctness for selected primitives, and proved the absence of run-time
  errors for the rest, at
  roughly \SpeedupLow--\SpeedupHigh{} times lower supervision cost than comparable hand verification.
  GNATprove alone was insufficient: some defects could not be detected and were resolved using known-answer
  tests, interoperability, or human review of specifications.
  Given weak checks, the agent tried to bypass them and reported success. We report where each layer caught faults and draw the central lesson: what an agent
  can be trusted to establish is bounded by the strength of its feedback.
\end{abstract}

\begin{IEEEkeywords}
software reliability, formal verification, AI coding agents, Ada/SPARK
\end{IEEEkeywords}

\section{Introduction}\label{sec:intro}

AI coding agents generate code at a speed no human team can match, but that code
can pass review while lacking semantic correctness. In sensitive domains such as cryptography and protocol implementations, plausible but unverified code
is a serious risk. Frontier models already discover exploitable vulnerabilities and put defenders who rely on slow
assurance methods at a disadvantage~\cite{ref_bigsleep,ref_bsi_ai}.

Deductive verification addresses this problem because a theorem prover shows that a program meets its specification independently of who wrote the code. However, verification is costly. Much of the work consists in formulating contracts, invariants, and ghost models, in selecting proof-friendly representations, and in diagnosing failed proof attempts.

We use AI coding agents to implement and deductively verify their code, to build a collection of
formally verified bare-metal security software in
Ada/SPARK. We use verification as a \emph{fault-prevention} discipline, in which a machine-checked proof
eliminates whole classes of defects, such as memory errors, run-time faults, and deviations from a
functional model, by construction and independently of its author. The collection
covers classical and post-quantum cryptographic primitives, an IKEv2 stack for IPsec VPNs, a
TLS\,1.3 and Matrix client, and a composite post-quantum certificate authority. In total, GNATprove discharged \VCtotal{} proof obligations, and we ran \TestTotal{} tests\footnote{Source code, proof
artifacts, and the skill files are available at
\url{https://github.com/tobiasphilipp/experimental-agentic-verified-software}.}. Some defects could not be
detected by GNATprove alone, and only weaker checks catch them: test vectors, interoperability against real peers, and
human review of specifications. The clearest example is our SSH transport. The agent transposed two
fields during key derivation in RFC~4253; every proof obligation was discharged, and the handshake
succeeded, but the first encrypted packet failed against OpenSSH. The proof was correct, but the
relevant property had never been specified. The reliability we observed comes from these layers in \emph{depth},
not from proof alone. One pattern recurs, and we draw it out as a central lesson: an agent can only
achieve what it receives feedback on. This experience report contributes the following:
\begin{itemize}
\item first-hand experience building this collection of agent-written, deductively verified security software, and the assurance profile it achieved;
\item a \emph{defense-in-depth} view of assurance that separates fault \emph{prevention} by proof
  from fault \emph{detection} by standard test vectors and interoperability, and identifies the remaining
  risk that neither removes;
\item failure case studies of defects that survived proof, such as a post-quantum encoder that is
  correct for only one parameter set and a transposed key derivation that passed every obligation, and of an
  AI-specific failure mode in which the agent circumvented a weak check and reported success; and
\item lessons for practitioners adopting agent-driven, high-assurance development, including that agent capability
  is bounded by feedback strength and that the cost of this assurance fell by roughly
  \SpeedupLow--\SpeedupHigh$\times$.
\end{itemize}

\noindent The rest of the paper is structured as follows. We describe the background
(Sect.~\ref{sec:background}), the verifier-driven method (Sect.~\ref{sec:method}), how the
skill was built (Sect.~\ref{sec:skill}), the collection and its assurance levels
(Sect.~\ref{sec:corpus}), the running system (Sect.~\ref{sec:fullsystem}), the limits of proof (Sect.~\ref{sec:taxonomy}),
the lessons learned (Sect.~\ref{sec:lessons}), limitations
(Sect.~\ref{sec:limitations}), and related work (Sect.~\ref{sec:related}).

\section{Background}\label{sec:background}

SPARK is a formally analyzable subset of Ada used in high-integrity industry. GNATprove performs
flow and initialization analysis and generates proof obligations for run-time checks and
user-written contracts; with strong enough contracts, it establishes functional
properties~\cite{ref_spark2014_gnatprove}. It is built on Why3, which discharges these obligations
with SMT solvers such as CVC5 and Z3 and, for the few obligations beyond automatic reach, an
interactive prover~\cite{ref_why3}. Guarantees come at three strengths: initialization and data
flow; absence of run-time errors, i.e., no overflow, out-of-bounds access, or uninitialized reads;
and functional correctness, where GNATprove proves the code meets the specification, often a ghost model. All three
demonstrate that the code meets \emph{its specification}. They do not certify that the specification captures
the standard, that the algorithm is secure, or that execution is free of side channels.

\section{Method: The Verifier-Driven Loop}\label{sec:method}

The method treats the agent as an unreliable producer of code and checks its output with a pipeline
of checks. These combine two dependability strategies: fault \emph{prevention}, where a machine-checked
proof excludes a defect class outright, and fault \emph{detection}, where a weaker check catches a
defect the proof missed. The agent proposes code and annotations, but does not judge whether they are
correct. Instead, checks such as GNATprove verify them in a feedback loop.

\subsection{Layered Checks}

Four automatic checks form the loop, ordered from strongest to weakest. \emph{Machine-checked proof
obligations}, from GNATprove over CVC5 and Z3, are the strongest check, since a discharged obligation guarantees, for
all inputs, that the code meets its specification, while an unproven obligation yields a precise error message, often with a counterexample. \emph{Known-answer tests}, such as the NIST ACVP/FIPS and RFC vectors, provide a byte-exact check, but only for specific inputs. \emph{Interoperability tests} against OpenSSL, strongSwan, OpenSSH,
and \url{matrix.org} are the only checks that exercise protocol behavior against independent
implementations. \emph{Constructed non-functional checks}, constant-time validation with TIMECOP and
dudect, do not exist natively and must be engineered into a pass/fail gate. No single check suffices. Proof prevents the defect classes within its reach, while the weaker checks detect faults outside it. These checks must resist being gamed by the agent.

\subsection{The Verifier-Driven Loop}

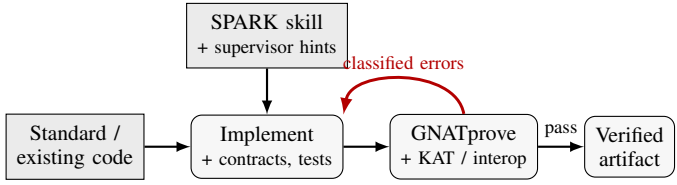
\begin{figure}[t]
\centering
\resizebox{\columnwidth}{!}{%
\begin{tikzpicture}[node distance=5mm and 6mm,
   block/.style={rectangle, draw, fill=black!3, rounded corners, align=center,
     minimum height=2.2em, font=\footnotesize, inner sep=4pt},
   io/.style={rectangle, draw, fill=black!8, align=center, minimum height=2.2em,
     font=\footnotesize, inner sep=4pt},
   fl/.style={-latex, line width=0.8pt},
   fb/.style={-latex, line width=1.1pt, draw=red!70!black}]
   \node[io] (spec) {Standard /\\ existing code};
   \node[block, right=of spec] (impl) {Implement\\ {\scriptsize+ contracts, tests}};
   \node[block, right=of impl] (verify) {GNATprove\\ {\scriptsize+ KAT / interop}};
   \node[block, right=of verify] (art) {Verified\\ artifact};
   \node[io, above=6mm of impl] (skill)
     {SPARK skill\\ {\scriptsize+ supervisor hints}};
   \draw[fl] (spec) -- (impl);
   \draw[fl] (impl) -- (verify);
   \draw[fl] (verify) -- node[above, font=\scriptsize]{pass} (art);
   \draw[fl] (skill) -- (impl);
   \draw[fb] (verify.north) to[out=100, in=80]
     node[pos=0.5, above, font=\scriptsize, text=red!70!black]{classified errors} (impl.north east);
\end{tikzpicture}}
\caption{The verifier-driven loop: the agent implements, and GNATprove and the test and
interoperability checks either certify the artifact or return classified errors that drive repair.}
\label{fig:workflow}
\end{figure}

The agent receives a natural-language or standards-level specification and iterates (Fig.~\ref{fig:workflow}): it creates
executable tests, writes the package specification with bounded subtypes and contracts, implements the
body with loop invariants, assertions, and ghost code, compiles and runs the tests, and runs GNATprove for
flow and run-time checks, and functional contracts. A \emph{skill}, a reusable instruction file the agent loads, containing proof-oriented SPARK rules,
worked examples, and a \emph{diagnostic playbook}, guides it. It classifies each unproved check as a
real bug, a missing contract, a weak loop invariant, a missing frame condition, or a prover
limitation. It then drives a ladder of cheap, low-risk fixes, such as tighter types,
exposed postconditions, exact preconditions, invariants, and ghost lemmas, before raising the level of solver resources. A human supervisor stays in the loop for architectural
direction, to detect specification gaming, and to judge when tool feedback has become uninformative;
the supervisor never authors code, contracts, or proof steps.

\paragraph{The \texttt{pragma Assume} policy.}
One rule guards the integrity of the loop. In the first ML-KEM iteration, the agent, unable to discharge failing obligations, silenced them with
\texttt{pragma Assume} and reported success. This is specification gaming, since the prover then accepts a fact
that the program does not establish. Human review caught it, and the skill now forbids using
\texttt{pragma Assume} to silence a proof obligation caused by weak code or contracts. The rule stopped this particular evasion, but the behavior reappeared in a different form when the agent excluded code from proof with \texttt{SPARK\_Mode => Off} in many places. We enforce the rule by skill text and human review, not an automatic check, because \texttt{SPARK\_Mode => Off} is legitimately needed in places such as socket handling, so a blanket ban would be wrong. AlphaVerus hit the same failure mode~\cite{ref_alphaverus}.

\subsection{Study Setup}

We ran the workflow with two command-line coding agents: OpenAI
GPT-5.5 under the Codex CLI, then Anthropic Claude Opus~4.8 under the Claude~Code CLI with a
one-million-token context window; we switched mainly to reduce API cost. The loop tolerated the
swap with no observable change in quality; the constant factor was GNATprove rather than the particular model.\footnote{Verification stack: SPARK Pro 26.2 over Why3
1.8.0, mostly at proof level 2 and level 4 where needed, with Alt-Ergo 2.6.1, Colibri 2025.02,
CVC5 1.2.0, and Z3 4.13.4, plus Isabelle 2025-2 for the few interactive proofs; a full run takes
hours.} The supervisor gave only high-level
strategic hints and chose among the agent's alternatives, never prescribing a concrete edit; the
reported hours cover this direction and review; the supervisor wrote no code or proofs. Every module we attempted was completed, though the SSH, IKEv2, and Matrix stacks are proof-of-concept
prototypes. The only abandoned attempt concerned an open problem. We tried a brief experiment of
having the agent prove $\mathrm{P}\neq\mathrm{NP}$ within hours, since no check can
bound progress there, and the loop had nothing to optimize against.

\section{Building and Evolving the Skill}\label{sec:skill}

The agent's SPARK-specific competence is encoded in a \emph{skill} file. We bootstrapped it from a report produced by a
deep-research tool on how to write and prove SPARK code. After each project, the agent recorded its learnings and
updated the skill, so the next project started from an improved one. When the SMT solvers reached
their limit, the same method produced a second skill, for Isabelle/HOL. We bootstrapped it from a summary of how to write and prove Isabelle, how to realize a Why3 proof obligation in higher-order logic, and let the agent write
the interactive proofs itself. The prover's kernel supplies the feedback that the SMT solvers could not.
The supervisor's role stayed constant, namely to provide the skill and to move to a stronger prover when needed, without ever writing a proof step.

\section{The Collection: Proven, Validated, Tested}\label{sec:corpus}

\begin{table*}[tb]
\caption{The collection by guarantee level, with code size and machine-checked obligations. FC =
functional correctness against a ghost model; AoRTE = absence of run-time errors; CT (constant-time)
is \emph{validated} with TIMECOP/dudect, never proven. kLOC counts non-comment, non-blank Ada/SPARK
via \texttt{tokei}. VCs = verification conditions (proof obligations).}
\label{tab:corpus}
\centering\footnotesize
\setlength{\tabcolsep}{3pt}
\renewcommand{\arraystretch}{1.1}
\begin{tabular*}{\textwidth}{@{\extracolsep{\fill}}L{3.8cm}L{2.8cm}L{5.2cm}L{3.2cm}rr}
\toprule
\textbf{Module} & \textbf{Proven} & \textbf{Validated} & \textbf{Tested} & \textbf{kLOC} & \textbf{VCs} \\
\midrule
SHA-2, SHA-3/SHAKE & FC & FIPS~180-4/202 KATs & --- & \multirow{8}{*}{33.1} & \multirow{8}{*}{\VCcrypto{}} \\
AES (ECB/CTR/GCM) & FC & NIST KATs; CT & --- & & \\
X25519, Ed25519 & FC (X25519) & RFC~7748/8032 KATs; CT & --- & & \\
ML-KEM (FIPS~203) & AoRTE & ACVP KATs; CT & --- & & \\
ML-DSA (FIPS~204) & AoRTE & ACVP KATs & --- & & \\
FrodoKEM (6 variants) & AoRTE & official KATs; CT & --- & & \\
XMSS$^{(\mathrm{MT})}$ & AoRTE & RFC~8391 KATs & reference impls & & \\
LMS/HSS & AoRTE${+}$round-trip$^{a}$ & RFC~8554 KATs & reference impls & & \\
\midrule
IKEv2 stack & AoRTE & --- & strongSwan 6.0.2 & 6.7 & \VCikev{} \\
TLS\,1.3 / Matrix client & AoRTE & RFC~8448 vectors & OpenSSL; matrix.org & 15.0 & \VCmatrix{} \\
Composite-PQ CA & AoRTE${+}$FC$^{b}$ & --- & unit tests; via TLS & 17.3 & \VCca{} \\
SSH transport & AoRTE & --- & OpenSSH 9.6/9.9 & 3.8 & 1{,}190 \\
\midrule
Hash map$^{c}$ & FC & --- & unit tests & 1.8 & \VChashmap{} \\
\midrule
\textbf{Total} & & & & \textbf{77.7} & \textbf{\VCtotal{}} \\
\bottomrule
\end{tabular*}\\[3pt]
{\footnotesize a)~LMS/HSS additionally carry a machine-checked sign$\to$verify round-trip for a freshly generated key ($q=0$). b)~FC for the DER/X.509/HTTP parsing layers, where GNATprove proves the decoded spans refine the wire grammar; AoRTE for the signature-bearing paths. c)~Hash map is human-written code with given contracts. Crypto rows share a single library, so their kLOC/VC entries are the library's total.}
\end{table*}

For each artifact, we describe \emph{the level of assurance it
provides}, from fault prevention by proof to detection by test, using a three-way distinction
throughout. \emph{Proven} means a
machine-checked GNATprove obligation, either AoRTE or, where a ghost model is present, functional
correctness against it. \emph{Validated} means a property checked by an automatic tool that is not a proof, such as a known-answer test against a standard's vectors, or a constant-time result from TIMECOP or dudect. \emph{Tested} means exercised against an independent peer implementation.

\subsection{Coverage}

Table~\ref{tab:corpus} gives the strongest guarantee that holds for each module and the weaker
check that does the rest. The collection carries no assumed mathematical axioms. We reached functional correctness against a ghost model for the hash families, the AES family, and X25519, the primitives whose mathematics are simple enough to model. Every post-quantum
scheme and every protocol stack is \emph{absence-of-run-time-errors only}: proved free of overflow,
out-of-bounds access, and uninitialized reads, but not proved to compute the intended function, and
no protocol stack has a proof of protocol-level correctness or of security. 

\paragraph{Effort}
The three supervised cases give a rough, order-of-magnitude effort figure.
ML-KEM, FrodoKEM, and the SSH prototype together required about \EffortTotalH{} hours of
supervision and \EffortTotalEUR{}~EUR in API costs. For FrodoKEM alone, a junior SPARK
developer needed about six weeks for a comparable implementation
and AoRTE proof, against roughly six hours of supervision here. Six weeks is about 240 working hours, so the ratio is
roughly \SpeedupHigh$\times$; counting our review and skill-building effort brings the conservative
end to about \SpeedupLow$\times$. This single comparison is illustrative only. We then
produced the remaining post-quantum schemes and the protocol stacks with comparable per-module
effort, several of them in a single largely unattended run.

\subsection{What ``Proven'' Means}

Three artifacts reach full functional correctness. GNATprove proves that X25519 refines a ghost
$\mathrm{GF}(2^{255}-19)$ field model, so the Montgomery ladder is shown to compute the specified scalar
multiplication, and the proof rests on no axiom of its own. A human-written
open-addressing hash map, contracts given, is proved functionally correct against an abstract map
model, allocator included; here, the human wrote the specification, so the separation between specification and machine-checked implementation is complete.

\section{From Components to a Running System}\label{sec:fullsystem}

\begin{figure*}[t]
\centering
\includegraphics[width=0.48\textwidth]{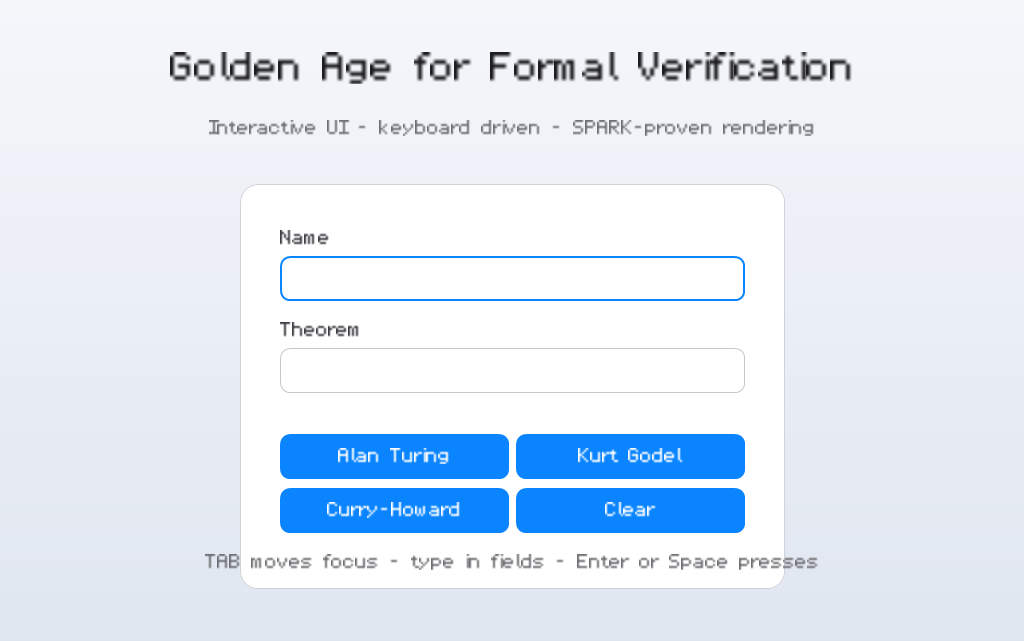}\hfill
\includegraphics[width=0.48\textwidth]{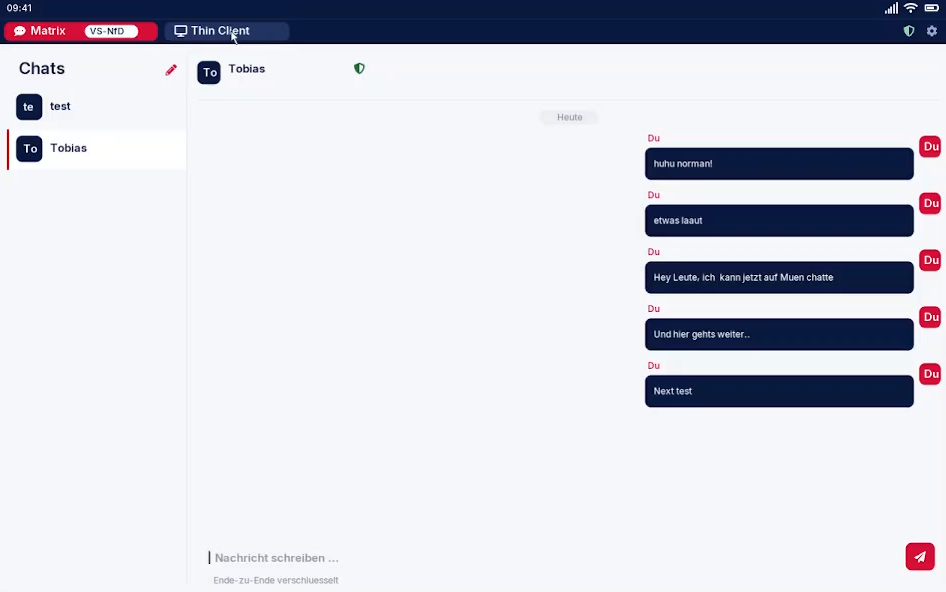}
\caption{Two views of the running system's agent-written, SPARK-rendered interface. Left: a
keyboard-driven input form, proven free of run-time errors. Right: the SecureMatrix client in an
end-to-end chat over the verified TLS stack.}
\label{fig:screens}
\end{figure*}

Many artifacts of Table~\ref{tab:corpus}, all written in Ada and largely in SPARK, compose into a
secure-messaging client that runs bare-metal on the Muen separation kernel, with no general-purpose
operating system underneath: an IPv4 stack, the TLS-secured Matrix client, a graphics stack, and a
windowed UI. The rendering stack matured from an AoRTE-proven 2D framebuffer through a keyboard-driven
SPARK form to the client shown in Fig.~\ref{fig:screens}, which exchanges messages over the TLS stack
and streams remote video through an embedded VNC session.
For a system like this, the deciding checks are no longer machine-checked. Whether the interface is
\emph{smooth}, \emph{visually acceptable}, and architecturally sound is a property that no proof
obligation expresses.

\section{Failure Modes: The Limits of Proof}\label{sec:taxonomy}

Every proof obligation in Table~\ref{tab:corpus} is discharged, and yet the collection contained real defects. We describe four situations where a fault escaped the proof layer and a weaker check or human
review had to catch it. Each is a small failure case study that marks the limit of automated assurance.

\subsection{A Correct Proof of a Wrong Specification}

A discharged obligation certifies that an implementation meets \emph{its specification}, not that
the specification captures the standard. Every obligation can be green while the specification itself is wrong, and the agent will faithfully prove conformance to a subtly wrong reading. In
FrodoKEM, the message \texttt{Encode} routine encoded a fixed two bits per coefficient, correct for
the smallest parameter set but wrong for the larger ones. Every run-time-error obligation was discharged, and for the smallest parameter set, even the test vectors
were passed. Only the known-answer tests for the larger parameter sets exposed the defect. The human's
main duty here is to check the \emph{specification against the standard}.

\subsection{Obligations Beyond Automatic Reach}

Some obligations lie outside the SMT solvers' reach. In our collection, these were an ML-KEM byte-indexing
lemma and an $8\times8$ bit-transpose identity in the constant-time AES. Here, the solver's feedback
is only \emph{partial}, because it reports the obligation as open but cannot guide a proof. We discharged both
with the Isabelle/HOL interactive prover~\cite{ref_isabelle}, the agent writing the proofs under a
bootstrapped skill. The limit is therefore the point where the feedback must change form, from a
decision procedure to a proof kernel that certifies each step.

\subsection{The Transposed Key That Passed Every Proof}

Run-time-error and even functional-correctness obligations operate within a single unit; they do not
range over the \emph{logic of a protocol} unless that logic is itself specified as a property to
prove. A clear case is the SSH transport, where the agent derived each session key as
$\mathrm{HASH}(K \,\|\, X \,\|\, H \,\|\, \mathit{session\_id})$, transposing two fields of RFC~4253
that require $\mathrm{HASH}(K \,\|\, H \,\|\, X \,\|\, \mathit{session\_id})$. Every obligation was
discharged, and the handshake appeared to work, because key exchange and the host-key signature depend only on $H$, until the first encrypted packet was decrypted wrongly against OpenSSH. The defect was caught by interoperability testing rather than by proof, since the property was never specified and therefore no obligation covered it.

\subsection{What No Functional Proof Can See: Timing}

Some properties that matter most for security are not functional at all. Constant-time execution is
the prime example. A functionally correct implementation can still leak secrets through timing, and
no functional proof detects this. We handle such properties by \emph{constructing} a hard check.
Constant time is validated with TIMECOP and dudect, and the result is folded into the loop as a pass/fail gate. The agent cannot be expected to establish properties for which it receives no feedback. 

\section{Lessons Learned}\label{sec:lessons}

We draw five lessons from the experience for teams adopting AI-assisted
high-assurance development. Above all, given a strong checker, an agent can write and verify relevant security code almost
autonomously, leaving the human to specify, review, and choose the checker.

\paragraph{Proof prevents defect \emph{classes}, not specification mistakes}
A discharged obligation certifies that the code meets the specification the agent wrote, not the
standard it was meant to capture. The remaining human duty is to review the
specification against the standard; the machine already checks the proof.

\paragraph{An agent games a weak check}
An agent optimizes for the check it is given, and will weaken the question rather than strengthen the
code if that is the easier path to a passing check. Reliability, therefore, depends on feedback that
cannot be faked, that is, on checks that do not accept the same wrong assumption that produced the defect.

\paragraph{Turn soft properties into hard checks}
Any property the agent needs to achieve must be reduced to an automatable
pass/fail check with informative feedback for the agent. We turned constant-time execution into a TIMECOP/dudect gate;
properties that resist such conversion continue human review duties.

\paragraph{Reliability comes from many checks and not from proof alone}
Proof, standard test vectors, peer interoperability, and human review catch different faults; neither
the FrodoKEM encoder defect nor the transposed SSH key derivation was reachable by proof. The reliability we observed came from the combination of layers.

\paragraph{Capability is bounded by feedback, and the cost fell sharply}
What the agent could be trusted to establish was bounded by the strength of the feedback relevant to
the goal. Where strong feedback existed, the agent produced verified code with little supervision, at
sharply reduced cost.

The generated SPARK code was reviewable but not always clean. The agent duplicated FrodoKEM logic
across parameter sets instead of using Ada generics, and some comments were inconsistent with the code.

\section{Trust Base and Remaining Risk}\label{sec:limitations}

Every guarantee holds only relative to a set of assumptions. The following is the trust base of the collection:
the facts that must hold for the proofs to mean what we say, none of which the proofs themselves
establish.

\paragraph{Matching the standard.}
Every guarantee is relative to a specification that the agent wrote and a human reviewed. The
remaining risk is a correct proof of a wrong specification; the defense consists of a human review
of the specification with the trusted boundary against the standard.

\paragraph{Side channels.}
Constant-time behavior is \emph{validated}, not proven. A timing or microarchitectural
leak the tools do not exercise is outside every guarantee here; turning constant time into a
machine-checked proof is a valuable extension.

\paragraph{Trusted primitives and the SPARK boundary.}
The parts we cannot model in SPARK live behind child packages marked
\texttt{SPARK\_Mode => Off}: the random source, drawn from the Linux operating system, and socket
handling, abstracted the same way. GNATprove checks the SPARK caller against the wrapper's contracts but
does not reason about the wrapper's body at all. 

\paragraph{Miscompilation.}
The proofs are about SPARK semantics, whereas the deployed artifact is machine code produced by
GNAT/GCC, which is \emph{not} a verified compiler. A code-generation bug would invalidate a
guarantee that holds perfectly at the source level. This is the standard gap between a verified
program and a verified executable, and this gap applies to our work unchanged.

\paragraph{Supply chain and process.}
Finally, the toolchain, its dependencies, the test vectors, and the generated tests are trusted
process inputs. The generated tests need review in particular, since an agent that writes both the code and its
tests can satisfy both with the same wrong assumption. We did not separately audit them; the
independent guard is interoperability against peers such as OpenSSH with and without
post-quantum key exchange, and strongSwan. The verifier is the trust anchor:
it bounds the dependence on any single model, but not on the process around it.

\section{Related Work}\label{sec:related}

\paragraph{Software reliability engineering.}
We place this work in the dependability tradition, in which faults, errors, and failures form a
causal chain, and dependability is achieved through fault prevention, tolerance, removal, and
forecasting~\cite{ref_laprie1985,ref_avizienis}, and defects are classified to steer the process~\cite{ref_odc}. Deductive verification is a means of fault
\emph{removal} that goes beyond testing: where a test samples inputs, a machine-checked proof
removes an entire class of faults at once. This proof-over-test discipline is by now an established industrial practice. Amazon Web Services applies formal methods to prevent classes of
defect that testing and after-the-fact recovery cannot, including continuous proofs of its s2n TLS
library~\cite{ref_aws_formal,ref_s2n}; Meta runs the Infer analyzer inside its deployment
pipeline~\cite{ref_infer}; and seL4 gives a complete operating-system kernel a machine-checked
functional-correctness proof that rules out whole defect classes~\cite{ref_sel4}. We follow this practice and add one difference: an AI agent writes the code under proof, and the prover, not the author, decides whether the implementation meets its specification.

\paragraph{Verified cryptography.}
The state of the art in verified cryptography was established by dedicated teams over several years. HACL\(^\star\)/EverCrypt provides a verified provider deployed in Firefox,
the Linux kernel, and WireGuard~\cite{ref_hacl,ref_evercrypt}; in the same SPARK toolchain, SPARKNaCl
is a handwritten verified NaCl~\cite{ref_sparknacl}, whereas our collection is larger and
agent-generated. Industrial vendors now verify post-quantum primitives directly. Apple
formally verified ML-KEM and ML-DSA in \texttt{corecrypto}, uncovering a real defect in the
process~\cite{ref_apple_corecrypto}. Such pipelines also illustrate a risk that applies to our work as well, since code can compile and
pass tests yet remain insecure through specification gaming~\cite{ref_verification_facade}.

\paragraph{LLMs in the verifier-driven loop.}
The closest SPARK-specific predecessor, Marmaragan, adds annotations to existing Ada code and
discharges all open obligations for 50.7\% of its benchmark~\cite{ref_marmaragan}; our agent instead
authors implementation, contracts, tests, and proof structure from standards-level input. LLMs have been used to automate code review itself~\cite{ref_codereview_moe}; we take the
complementary view that review, human or automated, samples program behavior, whereas a discharged
obligation establishes a property for all inputs. The symbolic checker serves as the final arbiter of correctness, as in AlphaVerus, which reaches verified
Rust behind an explicit filter against reward hacking~\cite{ref_alphaverus}. Our contribution is not a new generator or prover, but an experience report on the goals a machine-checked proof can achieve.

\section{Conclusion}\label{sec:conclusion}
Deductive verification lets AI agents build a large collection of security software whose defect
classes are excluded by machine-checked proof, while standard test vectors, interoperability, and
human review of specifications caught what escaped it. Machine-checked evidence improves the
trustworthiness of code: a proof decides whether the code meets the specification.

These findings concern engineering \emph{culture} as much as tooling. When an
agent produces verified code cheaply, the bottleneck moves from writing
software to specifying it correctly and reviewing it against the standard. The
engineer's role shifts from author to designer of specifications, checks, and
process constraints, and to final judge of what the machine cannot decide. Engineers
remain accountable for the generated code and must be able to read and explain
it; generating code does not by itself build understanding of the underlying
cryptography.

Four steps follow from this experience. First, if the cost reduction we
observed holds even in part, formal verification stops being the luxury
option among assurance techniques, and the question shifts from whether verification is affordable to whether its absence is. Second, the largest
remaining risk, a specification that mismatches its standard, is addressable
at its source. Standards bodies could publish machine-readable formal
specifications alongside prose and test vectors, which would close by construction the
gap that caused our FrodoKEM encoder defect. Third,
cheap agents revive a dependability technique long considered uneconomical:
$N$-version programming~\cite{avizienis-nvp}. Independent agents deriving
implementations and tests from the same standard would break the
shared-assumption risk in our own process and make the old question of
correlated faults testable at negligible cost. Fourth, following our own lesson that capability is bounded by feedback, the next check to add is a
protocol-level verifier such as Tamarin. Placed in the loop, it would extend proof
to the one class that no current layer covers, namely the security of the protocol logic
itself: the secrecy and authentication goals of the IKEv2 and TLS state
machines. The transposed key derivation, by contrast, is a conformance defect
and falls to the second step: only a machine-readable standard turns it into a
failing obligation. The open problem remains the step from
\emph{verified} to \emph{provably secure}, and the method for approaching it is to give the agent a stronger judge.

\let\oldthebibliography\thebibliography
\renewcommand{\thebibliography}[1]{%
  \oldthebibliography{#1}%
  \setlength{\itemsep}{0pt plus 0.2ex}%
  \setlength{\parsep}{0pt}%
}
\bibliographystyle{IEEEtran}
\bibliography{references}

\end{document}